\documentclass[aps,prl,10pt,notitlepage,nofootinbib,superscriptaddress,showkeys,showpacs, twocolumn]{revtex4-1}

\usepackage{amsmath,amssymb,amsthm,bbm}
\usepackage[english]{babel}
\usepackage{color}
\usepackage{graphicx}
\usepackage{subfigure}

\newcommand{\be}{\begin{equation}}
\newcommand{\ee}{\end{equation}}
\newcommand{\bqa}{\begin{eqnarray}}
\newcommand{\eqa}{\end{eqnarray}}
\newcommand{\bea}{\begin{eqnarray}}
\newcommand{\eea}{\end{eqnarray}}

\newcommand{\N}{\mathds{N}}

\DeclareMathOperator{\tr}{tr}

\def\N{{\mathbbm N}}

\begin{document}

\title{\Large \bf Multi-critical tensor models and hard dimers on spherical random lattices}

\author{{\bf Valentin Bonzom}}\email{vbonzom@perimeterinstitute.ca}
\affiliation{Perimeter Institute for Theoretical Physics, 31 Caroline St. N, ON N2L 2Y5, Waterloo, Canada}

\date{\small\today}

\begin{abstract}
\noindent Random tensor models which display multi-critical behaviors in a remarkably simple fashion are presented. They come with entropy exponents $\gamma = (m-1)/m$, similarly to multi-critical random branched polymers. Moreover, they are interpreted as models of hard dimers on a set of random lattices for the sphere in dimension three and higher. Dimers with their exclusion rules are generated by the different interactions between tensors, whose coupling constants are dimer activities. As an illustration, we describe one multi-critical point, which is interpreted as a transition between the dilute phase and a crystallized phase, though with negative activities.
\end{abstract}

\medskip

\keywords{Hard dimers, Random lattices, Random tensor models, continuum limit, critical behavior}

\maketitle

\section{Introduction}

Random matrix models \cite{mm-review-difrancesco} provide an efficient tool to understand some two-dimensional statistical field theories coupled to fluctuations of the geometry. They generate random discretizations of surfaces. Interesting features we want to stress here are $(i)$ multi-critical behaviors in the thermodynamic limit, $(ii)$ microscopic systems realizing them. By tuning some coupling constants, one achieves those multi-critical behaviors which are different universality classes corresponding to conformal field theories (CFT) coupled to 2d quantum gravity. 

It is a remarkable fact that it is often easier to solve such statistical mechanical models on dynamical lattices than on a fixed lattice, a fact usually credited to the reparametrization invariance which is gained on random lattices (see \cite{yang-lee-O(n)-kostov} for a recent application). A historical example is the 2d Ising model on planar graphs \cite{Kazakov:1986hu}, also exactly solvable in a magnetic field \cite{Boulatov:1986sb}, using a two-matrix model formulation which yields in the continuum the unitary Ising CFT (central charge $c=1/2$) coupled to quantum gravity. 

Other multi-critical points are observed in one-matrix models \cite{kazakov-matter}. The first one turns out to have the same entropy (or string susceptibility) exponent as the Ising model, which raised some confusion with the latter. But it was then shown that it can be derived from hard dimers on random lattices \cite{yang-lee-staudacher}. Hard dimers are objects which can be attached to links of a lattice with some exclusion rule. They can be obtained from the Ising system at infinite temperature in an imaginary magnetic field. Using a matrix model formulation, the partition function can be calculated exactly and the continuum limit is actually the Yang-Lee singularity, a non-unitary conformal field theory (CFT) with central charge $c=-22/5$ coupled to two-dimensional quantum gravity (see also \cite{ginsparg-moore-ising, gross-migdal-ising}). That universality class is found generically in models of hard objects \cite{hard-objects}. We want to emphasize that the derivation from a microscopic system (hard dimers) helped identify the continuum limit. The mechanism is that dimers generate an effective action for a one-matrix model with a new interaction potential. Hence, that provided generic one-matrix potentials with a physical picture, corresponding to non-unitary matter CFTs coupled to quantum gravity \cite{kazakov-matter, brezin-ising, graph-combinatorics-difrancesco, mm-review-difrancesco}.

To generate random lattices in dimensions $d>2$, one can use random rank-$d$ tensor models \cite{ambjorn-3d-tensors, sasakura-tensors, gross-tensors}. A class of such models (known as colored models \cite{color, Gur3, GurRiv, Gur4}) is now known to have a continuum limit in the large $N$ scaling limit ($N$ being the size of each tensor entry) \cite{Bonzom:2011zz}. While different universality classes have been found \cite{toy-double-scaling}, which are multi-critical points in the matrix model terminology, they $(i)$ come from a toy model which reduces to a rectangular-matrix model, $(ii)$ have not been physically motivated so far.

First guesses could be to run the Ising model or hard dimers, on those random lattices. However unlike in two dimensions,  using the large $N$ limit of a colored tensor model it has been found that there is no phase transition at finite temperature \cite{ising-colored} in agreement with previous numerical analysis \cite{ambjorn-3d-ising}.

In this article, we introduce a family of colored tensor models which exhibit multi-critical behaviors in a remarkably simple fashion. A way to characterize them is the entropy exponent, which quantifies the proliferation of microscopic configurations in the thermodynamic limit. We find the same entropy exponents as those of \cite{toy-double-scaling} which also corresponds to the entropy exponents of multi-critical branched polymers \cite{BP-ambjorn}.

Moreover, we provide a physical picture of such models. The presence of different interactions between tensors is interpreted in terms of dimer insertions on the dominant graphs, with some generalized exclusion rules. The coupling constants are the dimer activities. In the dilute phase, one recovers the entropy exponent of pure random lattices, strengthening its universality. Like in matrix models, multi-critical behaviors appear when activities are appropriately tuned. However, the thermodynamic limit is defined in a slightly different way from the standard one where it is strictly the number of lattice sites which is sent to infinity. We show that some multi-critical points correspond to phase transitions from the dilute phase  to a crystallized phase, though with negative activities. Finally, we discuss other interpretations of our model.

\section{Random tensors and hard dimers}

\subsection{Random tensors and random melonic triangulations of the sphere}

We start with the simplest colored tensor model which generates random triangulations of $d$-dimensional pseudo-manifolds. We take $d+1$ pairs of rank-$d$ tensors $(T^c_{n_1\dotsc n_d},\bar T^c_{n_1\dotsc n_d})_{c=0,\dotsc,d}$, each carrying a color index $c$, and having components $n_i=1,\dotsc,N$.  The tensor action has a quadratic part,
\be
S_{\rm link} = \frac1{g}\sum_{c=0}^d \sum_{\substack{
\left\{n_i=1,\dotsc,N, \right. \\
\left. {i=1,\dotsc,d}\right\}}}T^c_{n_1\dotsc n_d}\,\bar T^c_{n_1\dotsc n_d},
\ee
and an interaction part with a positive vertex and negative vertex,
\be
\begin{aligned}
S_+ &= \frac1{\sqrt{g}\,N^{\frac{d(d-1)}{4}}} \sum_{\{n_{ij}\}} \prod_{i=0}^d T^i_{n_{i i-1}\dotsc n_{i0} n_{id}\dotsc n_{ii+1}},\\
S_- &= \frac1{\sqrt{g}\,N^{\frac{d(d-1)}{4}}} \sum_{\{n_{ij}\}} \prod_{i=0}^d \bar T^i_{n_{i i-1}\dotsc n_{i0} n_{id}\dotsc n_{ii+1}}.
\end{aligned}
\ee
The coupling constant $g$ will be used to reach the continuum limit. 
The free energy
\be
e^{N^d F(g)} = \int \prod_{c=0}^d dT^c\, d\bar T^c\  e^{-S_{\rm link}-S_+ - S_-},
\ee
can be written as a sum over closed, connected, $(d+1)$-colored graphs,
\be
F(g) = \sum_{n\in\N} \sum_{\substack{\text{colored graphs}\\\text{$G_{2n}$ with $2n$ nodes}}} N^{-\omega(G_{2n})}\,g^{dn},
\ee
where $\omega(G)$ is a positive integer known as the degree of the graph \cite{Gur3, GurRiv, Gur4}. Those colored graphs are built from $(d+1)$-valent nodes of positive and negative orientations, respectively generated by $S_+$ and $S_-$. It means that the half-links around each node carry the colors $0$ to $d$, and are drawn on the plane by going clockwise or counter-clockwise. The propagator coming from $S_{\rm link}$ can only join a positive node to negative one, by connecting two half-links of the same color.

Those graphs encode triangulations of $d$-dimensional pseudo manifolds in the following way. First, we identify $k$-cells, $k=0,\dotsc,d$, on the graph as the connected, closed sub-graphs with exactly $k$ colors \cite{color}. Then, it is found that there is a colored triangulation which is dual to each graph, obtained by replacing each $(d-k)$-cell with a $k$-simplex. For example, a node is dual to a $d$-simplex and a link between two nodes corresponds to a $(d-1)$-simplex shared by two neighboring $d$-simplices. The boundary maps of the triangulations are defined by the boundary maps of the colored graph \cite{color}.

\begin{figure}
\includegraphics[scale=0.5]{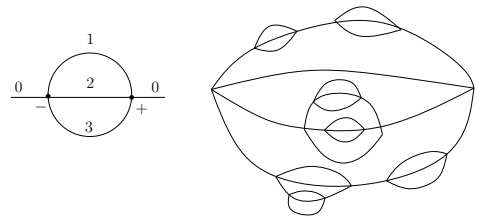}
\caption{\label{fig:melons} On the left, the elementary melon with external color 0. On the right a typical melonic graph, obtained by inserting melons inside melons.}
\end{figure}

In the large $N$ limit defined in \cite{Gur3, GurRiv, Gur4}, only triangulations of the $d$-sphere contribute. Moreover, they must have $\omega(G)=0$, from which they inherit specific combinatorics described in \cite{Bonzom:2011zz}, where they have been coined \emph{melonic} graphs. The basic building block is the \emph{elementary melon}, formed by two nodes of opposite orientations which are joined by $d$ links, as shown in the figure \ref{fig:melons}. The two external lines have the same color. The graphs are formed by gluing series of elementary melons next to each other, and inserting such sequences inside one another. They are melons inside melons so to say. To get a closed graph (hence a triangulation of the $d$-sphere) we join the two external lines. A typical graph is depicted in the figure \ref{fig:melons}, and we refer to \cite{Bonzom:2011zz} for details.

As a gluing of simplices, the elementary melon is built by gluing two $d$-simplices according to the rules, along $d$ of their boundary simplices. It has two external $(d-1)$-simplices of the same color which corresponds to the two external lines of the dual graph. If one thinks of the simplices as being flat and equilateral, then melonic triangulations carry strong positive curvatures.

The continuum limit is the regime where $F$ is dominated by graphs (triangulations) with a very large number of nodes ($d$-simplices). It is reached by sending the parameter $g$ to a critical value $g_c$, the radius of convergence of $F$ \cite{Bonzom:2011zz}. There $F$ becomes singular and behaves like $F\sim ( g_c-g)^{2-\gamma}$ with $\gamma=1/2$.

\subsection{Hard dimers}

\begin{figure}
\includegraphics[scale=0.4]{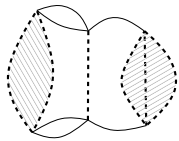}
\caption{ \label{fig:dimers} A graph dual to a triangulation at $d=3$ (since nodes are four-valent) with a set of two dimers on the left, a single dimer in the center and a set of three dimers on the right.}
\end{figure}

A dimer is an object which can be attached to a link of a lattice. We consider hard dimers, with a generalized exclusion principle. It is usually considered that two dimers can never touch each other. Here we choose that a node can share dimers with only one neighboring node. It is reasonable if we think of a dimer as a bond between two monomers. If two nodes with monomers are joined by several links, they can form a bond on each link, provided they do not share a dimer with another partner. Interpreting further, two dimers joining the same two nodes define a surface whose boundary is formed by the dimers and along which the dimers may couple to each other. If a third dimer is added, one gets three surfaces which glue together to form a closed surface. It is bounding a 3-cell which can be interpreted as a three-dimensional coupling region. Such coupling regions can have dimension $d$ at most. An example is given in the figure \ref{fig:dimers}. An activity $z_k$ is assigned to each set of $k$ dimers joining two nodes, and the partition function is
\be
\Xi_{G_n}(z_k,g) = \sum_{D\in\mathcal D(G_n)} g^{-\vert D\vert} \prod_k z_k^{\vert D_k\vert}.
\ee
Here $D$ is a dimer configuration, i.e. an assignment of dimers to the lattice satisfying our exclusion principle, and $\mathcal D(G_n)$ is the set of dimer configurations. $\vert D_k\vert$ is the number of sets of $k$ dimers joining two nodes in $D$ and we have set $\vert D\vert =\sum_k k \vert D_k\vert$ the total number of dimers in $D$. At this stage $g$ is just a book-keeping parameter which can be absorbed into a redefinition of the activities.

\begin{figure}
\includegraphics[scale=0.5]{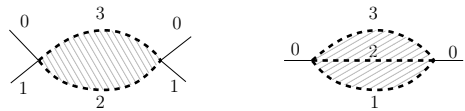}
\caption{\label{fig:2-dimer} The dimer parts of the action can be drawn with internal (dashed) lines which represent dimers and internal faces (2 lines and 1 face for the set of two dimers on the left, 3 lines and 3 faces for the set of three dimers on the right). The effective interaction only sees the external legs carrying some $T$ and $\bar T$.}
\end{figure}

Making the lattice dynamical means that we sum over a set $\{G_n\}$ of graphs. If colored graphs are considered, that can be implemented using the above tensor model supplemented with effective interactions which create hard dimers. Links without dimers are created by $S_{\rm link}$, positive and negative nodes without dimers by $S_\pm$. A set of $d+1-k$ dimers with colors $\{0,\dotsc,d\}\setminus\{c_1,\dotsc,c_{k}\}$ between two nodes is obtained by connecting a vertex $+$ and a vertex $-$ with links of colors $\{0,\dotsc,d\}\setminus\{c_1,\dotsc,c_{k}\}$. The external lines have colors $\{c_1,\dotsc,c_{k}\}$ and carry indices determined by the form of $S_+,S_-$. The scaling with $N$ is also determined by $S_+,S_-$ and by the $\frac{(d+1-k)(d-k)}{2}$ internal faces formed by the dimers. That leads to effective interactions, i.e. from the view of the external lines which do not carry dimers, of the form
\be
S^{\{c_1,\dotsc,c_{k}\}}_{\rm dimers} = - \frac{z_{d+1-k}}{g\ N^{\frac{(k-1)(2d-k)}{2}}}\ \tr\, \prod_{i=1}^k T^{c_i} \bar T^{c_i}.
\ee
We have denoted the index contractions with a trace, as it is otherwise a bit cumbersome. We have depicted $S^{\{0,1\}}_{\rm dimers}$ and $S^{\{0\}}_{\rm dimers}$ for $d=3$ on the figure \ref{fig:2-dimer}. The terms with $k=1$ are just corrections to $S_{\rm link}$.

The free energy $\Xi(z_k,g)$ is defined by
\be
\begin{aligned}
&e^{N^d \Xi(z_k,g)} = \int \prod_{c=0}^d dT^c\, d\bar T^c\  e^{-S},\\
&S = S_{\rm link}+S_+ +S_- + \sum_{k=1}^{d-1} \sum_{\{c_1,\dotsc,c_k\}} S^{\{c_1,\dotsc,c_k\}}_{\rm dimers}.
\end{aligned}
\ee
For simplicity\footnote{It is possible to solve the model similarly with $z_1\neq 0$. However, this becomes more complicated and does not bring different qualitative results.}, we set $z_1=0$. $S_{\rm dimers}$ does not affect melonicity of the large $N$ limit but instead really introduces dimers on melons. A typical Feynman graph is a closed connected melon $G_{2n}$ with a dimer configuration $D$ and the free energy has the expansion
\be
\Xi(z_k,g) = \sum_{n\in\N} g^{dn} \sum_{\{G_{2n}\}} \Xi_{G_{2n}}(z_k,g).
\ee
It is interpreted as a grand-canonical partition function for dimers, where one sums over all possible melonic triangulations and all dimer configurations. There is an associated canonical partition function where the quantity $p=dn-\vert D\vert$ is kept fixed,
\be
\Xi_p(z_k) = \sum_{n\in\N} \sum_{\{G_{2n}\}} \sum_{\substack{D\in\mathcal{D}(G_{2n}),\\ \vert D\vert = dn -p}} \prod_k z_k^{\vert D_k\vert},
\ee 
such that 
\be
\Xi(z_k,g) = \sum_{p\in\N} g^p\ \Xi_p(z_k).
\ee
The corresponding thermodynamic limit is the large $p$ regime. Note that it is not the standard regime of lattices with an infinite number of sites (studied in \cite{yang-lee-staudacher} for instance), as $p=dn-\vert D\vert$ also depends on the number of dimers. Equivalently, $\ln g$ is not a chemical potential for the number of lattices sites $n$, but for $p$ instead. The large $n$ regime will be studied in a companion paper, to appear.

To make sense, $\Xi_p(z_k)$ should behave like
\be
\Xi_p(z_k) \sim A\ p^{\gamma-3}\ g_c(z_k)^{-p},
\ee
for some constant $A$ and some function $g_c(z_k)$. The exponential decay is expected since $[g_c(z_k)]$ is then the radius of convergence of $\Xi(z_k,g)$. For a fixed set of activities $\{z_k\}$, when $g$ is tuned to $g_c(z_k)$, the grand-canonical partition function $\Xi(z,g)$ starts to be dominated by lattices with very large values of $p$ (again, this is slightly different from being dominated by lattices with a very large number of sites). The power-law decay $p^{\gamma-3}$ then determines the singular behavior of the system close to that critical surface,
\be \label{def gamma}
\Xi(z_k,g) \sim (g-g_c(z_k))^{2-\gamma}.
\ee
$\gamma$ is known as the entropy exponent, as it is related to the way the microscopic configurations involved in $\Xi_p$ proliferate at large $p$ (also known in two dimensions as the string susceptibility exponent, for which one-matrix models give $\gamma=-1/m$).

\section{Thermodynamic limit and multi-critical behaviors}

A key object of the analysis is the full, connected 2-point function which we write
\be
\langle T^c_{n_0\dotsc n_d}\bar T^c_{n'_0\dotsc n'_d} \rangle = U(z_k,g)\ \prod_i \delta_{n_i,n'_i}.
\ee
Then,
\begin{multline}
g\frac{\partial F}{\partial g} = \frac{d+1}{g}\,U + \frac{\langle S_+ + S_-\rangle}{2 N^{d}} \\
+ \sum_{k=1}^{d-1} \sum_{\{c_1,\dotsc,c_k\}} \frac{\langle S^{\{c_1,\dotsc,c_k\}}_{\rm dimers}\rangle}{N^d}.
\end{multline}
In the melonic sector, all the above expectation values are monomials in $U$. Part of the proof can be found in \cite{universality-gurau}, and more details will be appear soon elsewhere. In \cite{Bonzom:2011zz}, it was shown that $\langle S_+\rangle\sim\langle S_-\rangle \sim U^{d+1}$. Moreover, $\langle S_{\rm dimers}^{\{c_1,\dotsc,c_k\}}\rangle$ is obtained by reconnecting the external $T^{c_i}$ with the external $\bar T^{c_i}$ using the full 2-point function, which gives a $U^k$ behavior. Hence if $F$ has a singular part like \eqref{def gamma}, the most singular terms in the above equation are those going like $U$, with $U\sim (g_c(z_k)-g)^{1-\gamma}$.

\begin{figure}
\includegraphics[scale=0.4]{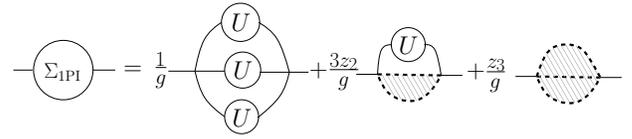}
\caption{\label{fig:eq-melons-dimers} In the melonic sector the 1PI 2-point function is built from the full 2-point function $U$ and dimer insertions. The first term of the right hand side is that for pure random lattices \cite{Bonzom:2011zz}, due to $S_\pm$, and the others are due to effective sets of dimers between the two external lines. }
\end{figure}

$U$ can be evaluated as the geometric series of the 1-particle irreducible 2-point function $\Sigma$, $U=g/(1-g\Sigma)$, say with external color 0. A self-consistency equation for $U$ is obtained if one can express $\Sigma(U)$ independently. It is easy in the melonic sector. One contribution comes from the dimer-free nodes, $S_\pm$. We glue a vertex $+$ with a vertex $-$ by inserting $d$ full 2-point functions, one on each color $c=1,\dotsc,d$. The terms $S_{\rm dimers}$ also contribute. One isolates the two lines of color 0 as the external lines, and connect the other external lines which have the same color inserting $U$. That gives
\be
\Sigma = \frac1{g}\bigl[ U^d +\sum_{k=0}^{d-2} z_{d-k} \begin{pmatrix} d\\k\end{pmatrix} U^k\bigr],
\ee
also symbolically depicted in the figure \ref{fig:eq-melons-dimers} for $d=3$. The closed equation on $U$ and $g$ is
\be \label{g(U)}
g = (1-z_d)U - \sum_{k=1}^{d-2} z_{d-k}\begin{pmatrix} d\\k\end{pmatrix} U^{k+1} - U^{d+1}.
\ee

By definition, $U$ is a series in $g$ and the thermodynamic limit corresponds to the regime where $g$ goes to the radius of convergence of $U$. The criterion for that loss of analyticity is
\be \label{dg/du}
\frac{\partial g}{\partial U} =0,
\ee
which is polynomial in $U$ of degree $d$ (in particular, it has explicit solutions for $d=3,4$). When the activities are small enough, the physical root is that passing through\footnote{This is the model studied in \cite{Bonzom:2011zz} though with different conventions.} $U_c(z_k=0)=1/(d+1)^{1/d}$, and satisfying
\be \label{pure lattice}
1-z_d = (d+1) U^d + \sum_{k=1}^{d-2} z_{d-k}(k+1)\begin{pmatrix} d\\k\end{pmatrix} U^{k}.
\ee
That gives a critical surface $z_d(z_2,\dotsc,z_{d-1},U)$. Starting from the physical solution $U(z_k=0)$, it can be inverted to give $U(z_k)$, hence $g(z_k)$, as long as $\partial z_d/\partial U\neq0$. Along this surface $\gamma=1/2$ and it shows the universality of the critical behavior of pure random melonic lattices observed in \cite{Bonzom:2011zz}.

The condition $\partial z_d/\partial U=0$ is
\be
-2d z_{d-1} = (d+1)d U^{d-1}+ \sum_{k=2}^{d-2} z_{d-k}(k+1)k\begin{pmatrix} d\\k\end{pmatrix} U^{k-1},
\ee
which restricted to the critical surface \eqref{pure lattice} gives one relation between the activities. Since $\partial z_d/\partial U=0$ is equivalent to $\partial^2 g/\partial U^2=0$, one has $(U-U_0)\sim (g-g_c(z_k))^{\frac{1}{3}}$ and hence $\gamma=2/3$ on this sub-surface.

It actually holds as long as $\partial z_{d-1}/\partial U\neq 0$. When the derivative is zero, we get a second relation between the activities which corresponds to $\partial^3 g/\partial U^3=0$, so that $\gamma=3/4$. This reasoning can be repeated so that at each step the activities $z_d,\dotsc, z_m$ can be expressed in terms of $z_{m-1},\dotsc,z_2$.

As $\partial g/\partial U$ is a polynomial, that set of relations means that some of its roots meet. When the activities are varied, the physical solution may intersect with other roots and jump from the universality class of pure random lattices to a new one. If exactly $m=2,\dotsc,d-1$ roots meet,
\be
\frac{\partial^k g}{\partial U^k}=0,\ k=1,\dotsc,m,\quad \text{but}\quad\frac{\partial^{m+1}g}{\partial U^{m+1}}\neq 0.
\ee
Those relations determine $m-1$ activities as functions of the others. On such points $U\sim (g-g_c)^{1/(m+1)}$, so that multi-critical points with
\be
\gamma = \frac{m-1}{m},\ \text{for}\quad m=2,\dotsc,d,
\ee
are obtained. Like in two dimensions \cite{yang-lee-staudacher}, that requires some activities to be negative, hence ``non-physical''.

The behavior of the system may be complicated in the neighborhood of such points. There are two notable situations we want to mention.
$(i)$ The \emph{cut singularity}, when it is not possible to go beyond some values of the parameters. A well-known example in two dimensions is the Yang-Lee singularity which is indeed the tri-critical point of the matrix model for hard dimers \cite{yang-lee-staudacher}, and also appears generically in models of hard objects \cite{hard-objects}. Technically, it is due to the collapse of the physical branch with another root of $\partial g/\partial U$ such that both solutions become complex beyond the singularity.
$(ii)$ The \emph{phase transition}, when the physical branch meet another root of $\partial g/\partial U$ and that they exchange their role after the critical point. A famous example is the 2d Ising transition on random lattices at finite temperature observed in a two-matrix model \cite{Kazakov:1986hu} and also in models of hard objects on bicolorable random lattices \cite{hard-objects}.

\section{Phase transitions}

Note that the coefficients of the polynomial $\partial g/\partial U$ can be chosen arbitrarily except for the vanishing of the term $U^{d-1}$ which corresponds to the constraint that the sum of the roots is zero. That leaves enough freedom to observe the typical behaviors mentioned above. While it is possible to do it with $d=3$ or $4$, it is easier and more natural to go to higher dimensions\footnote{To observe a phase transition with $d=4$, one has to increase $z_4$ above $1$, which means that the quadratic term in the action has a vanishing coupling at some point and then comes with the wrong sign. While it seems there is no trouble with the self-consistency equation, we do not know to which extent going through $z_4=1$ is problematic.}. In the remaining we consider $d=6$ with $z_1=z_3=z_5=0$, so that
\be
\frac{\partial g}{\partial U}= -7U^6 - 75 z_2U^4 - 45 z_4 U^2 + 1 - z_6.
\ee
For small activities, it is a phase of random lattices with dilute dimers. Therefore it can be described perturbatively, and $U$ grows linearly with the activities. For large activities, we expect a phase of fully packed dimers. The main contributions then come from $S_{\rm dimers}$, which means that the term $-7U^6$, coming from nodes without dimers, can be neglected. Assuming the dimer terms all have the same order, one finds typically that $U/\sqrt{z_2}$ goes to zero. It is actually expected that the free energy does not change too much with $z_k$ in the crystallized phase.

When all activities are zero, $U=(1/7)^{1/6}$. First, we lower $z_6$ from $0$ to $-6$, so that $U=1$. The equation \eqref{g(U)} for $g(U)$ has two real extrema, the physical continuum limit being the positive one, and correspondingly $\partial g/\partial U$ has to real roots, $\pm 1$, and four complex roots $e^{2ik\pi/6}$, $k=1,2,4,5$. To observe multi-criticality, it is necessary to create two real positive roots, and to collapse one of them with the physical root. Therefore we tune some activities to create two real positive roots, setting $z_2=-\frac{21}{25}, z_4 = \frac{56}{15}$ and $z_6=-111$ for instance.

Then we find a path to collapse and exchange the role of the physical root with another one. For example,
\begin{align}
z_4(z_2) &= -\frac{91}{60}+\frac{5}{28} z_2(25 z_2-14),\\
z_6(z_2) &= -174 - \frac{375}{7}z_2(15 z_2+14).
\end{align}
The real positive roots on this path are
\begin{align}
U_0(z_2) &= -\frac12 +\frac32 \sqrt{-\frac{50}{21}z_2-1},\\
U_1(z_2) &= 2,\\
U_2(z_2) &= \frac12 +\frac32 \sqrt{-\frac{50}{21}z_2-1}.
\end{align}
For $-\frac{119}{75}< z_2 < -\frac{21}{25}$, one has $U_0<U_1<U_2$ and the physical solution is $U_0$. However, $U_0$ and $U_1$ meet when $z_2$ reaches the critical value $z^*_2=-\frac{119}{75}$ and $\gamma=2/3$ at this point. For $z_2<z_2^*$, $U_1<U_0<U_2$.

To see there is a phase transition, notice that $U_0\sim \sqrt{-z_2}$ at large negative activities, so that it is not a physically reasonable solution. By contrast, $U_1^6\ll z_2 U_1^4$, which means that the constant solution is meaningful when nodes with dimers dominate. We conclude that $U_1$ is the solution for $z_2<z_2^*$ and that there is a crystallization transition between a dilute phase and a phase with fully packed dimers.

\begin{figure}
\subfigure[\ $U(z_2)$ with a discontinuity in $\frac{dU}{dz_2}$ at $z_2^*$]{\includegraphics[scale=0.4]{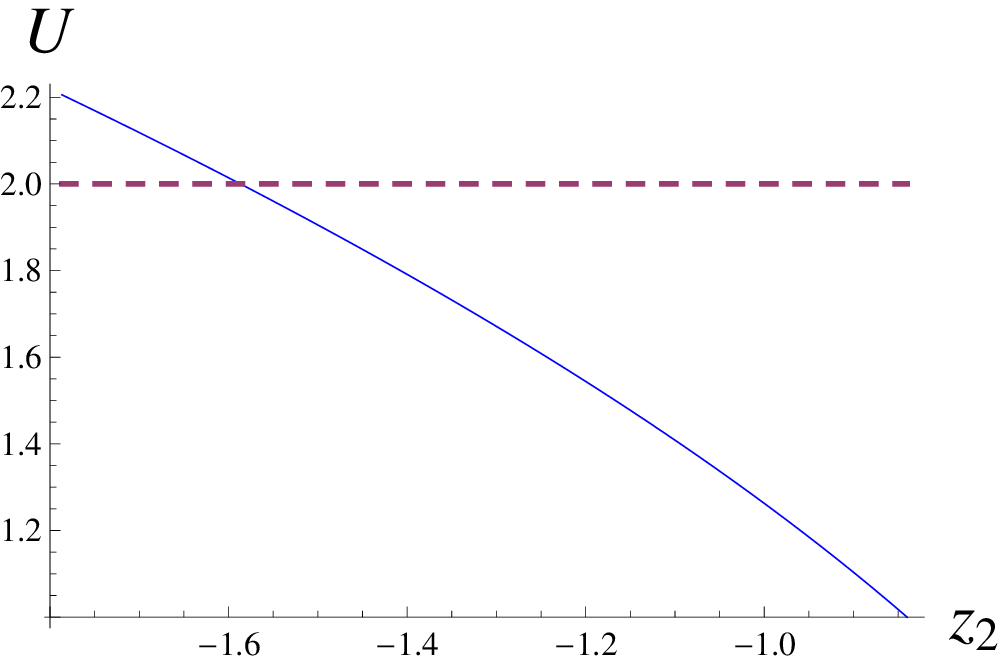} \label{fig:U-z2_2ndorder}} \
\subfigure[\ $U(z_2)$ with a discontinuity in $\frac{d^3U}{dz_2^3}$ at $z_2^*$]{\includegraphics[scale=0.4]{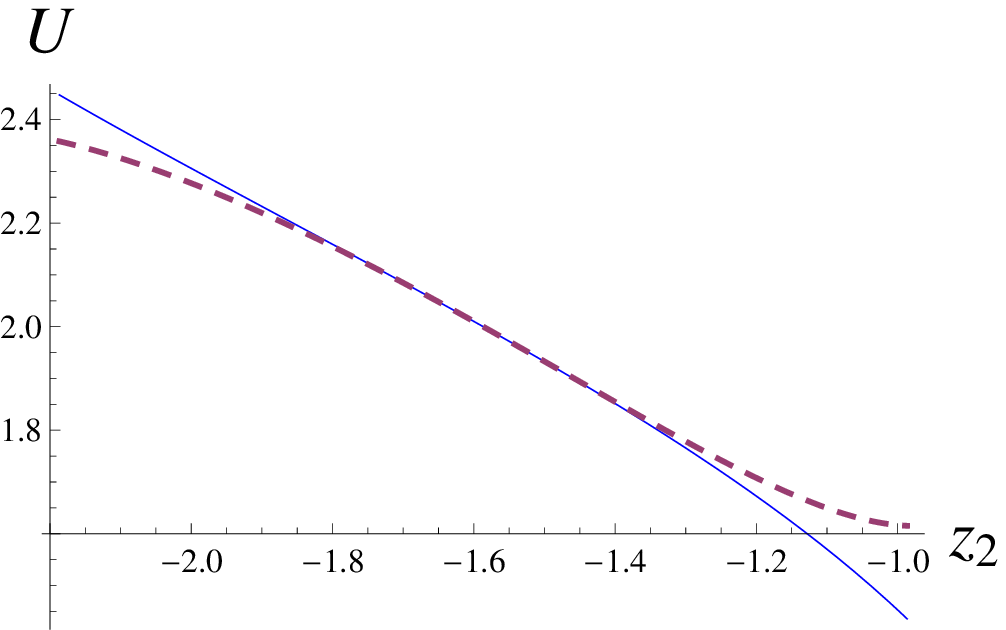} \label{fig:U-z2_4thorder}}
\caption{Plots of $U(z_2)$ for different paths to the phase transition. Before the transition, $z_2>z_2^*$, the physical solution is $U_0$ (normal style), and for $z_2<z_2^*$, it is $U_1$ (dashed).}

\end{figure}

In the continuum limit the free energy is a function of $z_2$, $F(z_2, z_4(z_2),z_6(z_2),g(z_2))$. The order of the transition is determined by the discontinuity in some derivative of $F$ with respect to $z_2$. The same way $\partial F/\partial g$ can be written in term of $U$, $dF/dz_2$ is also a polynomial in $U$. According to the path we have chosen, $U$ is obviously continuous at the transition (see figure \ref{fig:U-z2_2ndorder}), but its derivative is not, and therefore the transition is second order.

However, it is possible to choose other paths in the space of activities to make the transition as smooth as one wants.
Let us give an example. Take a parametrization such that $U_0=2+x$ for $z_2>z_2^*$, the transition being at $x=0$. Instead of setting $U_1$ as a constant like above, one can make it behave like $U_1\simeq 2+x-x^3$ around $x=0$. For definiteness, choose $U_2=3+x$ and the negative roots as the opposite of the positive ones. One can expand such a polynomial to identify $z_2(x), z_4(x), z_6(x)$. $z_2$ is of order 6 in $x$ but we can use some truncation to extract $x(z_2)$ in the neighborhood of $z_2=-119/75$. Since the change of variables between $x$ and $z_2$ is well-defined, one concludes that only $d^3U/dz_2^3$ is discontinuous at the transition, so that the transition is of the fourth order (see figure \ref{fig:U-z2_4thorder}).

\section{Applications}

\subsection{Monomers on random branched polymers}

The system of dimers on random melonic triangulations can be mapped to a problem of monomers on random branched polymers. That is suggested by the self-consistency equation \eqref{g(U)} which is similar to the algebraic equations on trees which are also found in matrix models \cite{graph-combinatorics-difrancesco}. First, there is a one-to-one correspondence between melonic triangulations and a family of branched polymers, due to a partial order on the set of melons in a melonic graph \cite{Bonzom:2011zz}. Those branched polymers are rooted $(d+1)$-ary trees, as any 1PI connected 2-point subgraph of the graph dual to a triangulation is represented by a vertex of degree $(d+2)$. A vertex of the tree either is a leaf (it has no child) or it has $d+1$ descendants.

Dimers with their exclusion rules are simply mapped to monomers attached to branched polymer vertices which are \emph{not} leaves. We have $d-1$ types of monomers, coming with activities\footnote{We again assume for simplicity $z_1=0$.} $z_2,\dotsc,z_d$. While the use of monomers on branched polymers trivializes the dimer exclusion rules, there is one subtlety which actually comes from the coloring. Remember that the dual graph to a triangulation has colors on its links. They become colors on the links of the corresponding tree. A vertex comes from its parent with a link of color, say, $i$, and its descendants are joined with links of colors $0$ to $d$, including the color $i$. The rule is that such a vertex can carry one monomer with activity $z_2, z_3,\dotsc$ or $z_k$ provided it has at least $k$ children which are leaves with colors different of $i$.

Intuitively, one can think of such monomers as being able to grasp some leaves attached to a vertex, but being sensitive to the color of their parent, they cannot grasp a leaf with the same color.

\subsection{Two-dimensional interpretation}

Finally, we would like to draw attention towards a different interpretation of our model. The triangulations we built are obviously $d$-dimensional. However, their dual graphs in the melonic sector are \emph{planar} graphs. More precisely, they form a \emph{subset} of the planar graphs encountered in matrix models. This suggests a interpretation on the 2-sphere, but we are not able to describe it yet. Remarkably, going to this subset of planar graphs changes the universality classes observed in 2d gravity coupled to matter. The 2d entropy exponents are of the form $\gamma_{2d} = -1/m$ (with $m=2$ for pure random lattices, i.e. pure gravity) while we got families of the form $\gamma = 1+\gamma_{2d}$ ($m=2$ being again pure random lattices \cite{Bonzom:2011zz}). Technically, one can count planar graphs with two external legs not necessarily in the same face using decorated trees \cite{graph-combinatorics-difrancesco}, resulting in the same asymptotic behavior as for melons. But the free energy is not directly related to it by a simple derivative.

It is known that matrix models at multi-critical points correspond to conformal field theories coupled to 2d quantum gravity, and it would be interesting to find whether the same statistical models on melons also take part to the conformal field theory landscape. A first hint would be to rewrite the Schwinger-Dyson equations derived in \cite{Gurau:2011tj} as Virasoro constraints.

\section*{Acknowledgements}

The author is indebted to Razvan Gurau, for numerous discussions and his active support during the early stage of this work.

Research at Perimeter Institute is supported by the Government of Canada through Industry Canada and by the Province of Ontario through the Ministry of Research and Innovation.


\end{document}